\documentclass[a4paper,fleqn]{cas-sc}

\usepackage[authoryear,longnamesfirst]{natbib}
\usepackage{graphicx}
\usepackage{amssymb}

\usepackage{amsmath}

\usepackage{listings}
\usepackage[protrusion=true,expansion=false]{microtype}
\usepackage{threeparttable} 

\def\tsc#1{\csdef{#1}{\textsc{\lowercase{#1}}\xspace}}
\tsc{WGM}
\tsc{QE}
\setcitestyle{numbers,square}

\begin{document}
\renewcommand{\bibfont}{\fontsize{8pt}{9pt}\selectfont}
\let\WriteBookmarks\relax
\def\floatpagepagefraction{1}
\def\textpagefraction{.001}

\shorttitle{AngstromPro}    

\shortauthors{H. Zhao et al.}  

\title [mode = title]{\textit{AngstromPro}: A software platform for STM data management, visualization and analysis}  



\author[a,b]{Huiyu Zhao}
\fnmark[1]
\credit{Conceptualization, Methodology, Software, Validation, Formal analysis, Writing -- original draft}

\author[c]{Jiahao Yan}%
\fnmark[1]
\cormark[1]
\ead{jhyan@ucc.ie}
\credit{Conceptualization, Methodology, Software, Validation, Investigation, Writing -- review \& editing, Supervision}

\author[c]{Catherine Dawson}
\credit{Data curation, Writing -- review \& editing}

\author[a,b]{Haitao Yang\corref{cor2}}
\cormark[1]
\ead{htyang@iphy.ac.cn}
\credit{Conceptualization, Methodology, Writing -- review \& editing, Supervision}

\author[a,b,d]{Hong-Jun Gao}
\credit{Conceptualization, Writing -- review \& editing, Funding acquisition}

\affiliation[a]{organization={Beijing National Center for Condensed Matter Physics and Institute of Physics, Chinese Academy of Sciences}, addressline={Beijing 100190}, country={China}}
\affiliation[b]{organization={School of Physical Sciences, University of Chinese Academy of Sciences}, addressline={Beijing 100190}, country={China}}
\affiliation[c]{organization={School of Physics, University College Cork}, city={Cork}, country={Ireland}}
\affiliation[d]{organization={Hefei National Laboratory}, addressline={Hefei 230088}, state={Anhui}, country={China}}
\cortext[1]{Corresponding author}
\fntext[1]{These authors contributed equally to the work.}


\begin{abstract}
Modern scanning tunneling microscopy (STM) experiments increasingly generate multidimensional datasets that require coordinated data management, visualization, and analysis. We present AngstromPro, an open-source Python-based software platform designed to support these activities within a common interactive environment. Its application architecture organizes in-application data through independent module instances, each combining a Workspace with a common structure and module-specific visualization and interaction. Scientific processing methods are registered independently of individual modules and operate on explicitly selected inputs through common execution services. A unified I/O subsystem maps heterogeneous on-disk formats to supported in-application data types. The plugin subsystem can add new modules, processing methods, I/O handlers, and in-application data types without modifying the core package. Representative workflows demonstrate the integration of established methods. In the registration example, the normalized cross-correlation maximum increases from 0.855 to 0.966. AngstromPro therefore provides an integrated software architecture for organizing, visualizing, processing, and extending STM data-analysis workflows.
\end{abstract}


\begin{highlights}
\item Open-source platform integrating STM data management, visualization, and analysis.
\item Module-based Workspaces provide independent contexts for STM analysis.
\item Processes are registered independently of modules for use with compatible inputs.
\item Unified I/O and plugins support heterogeneous file formats and platform extension.
\item Workflows demonstrate the platform, with quantitative registration validation.
\end{highlights}


\maketitle

\section{Introduction}
Since the invention of the scanning tunneling microscope (STM) by Binnig et al.\cite{Binnig1982}, STM has become an important tool in materials science and nanotechnology. Modern STM systems can operate with ultra-low vibration, under ultra-high vacuum, at ultra-low temperatures, and in high magnetic fields\cite{Fein1987, Pan1999, Song2010, Misra2013, Singh2013, Assig2013, Tao2017, Esat2021, Gong2021}. With lock-in detection, scanning tunneling spectroscopy (STS) provides spatially resolved measurements of the local electronic density of states with energy resolution reaching the micro-electron-volt scale. Beyond conventional STM and STS, a range of instrumental developments has further expanded the information accessible by scanning tunneling microscopy, including radio-frequency spin resonance\cite{Friedlein2019, Drost2022, Hwang2022}, cryogen-free refrigeration\cite{DenHaan2014, Kasai2022, Ma2023}, quartz-resonator probes\cite{Jung2017, Cheng2022}, fast scanning\cite{Yang2022}, electrical-transport characterization\cite{Ma2017, Liebmann2017, Schwenk2020, Yan2021}, spin-polarized measurements\cite{Wang2011, VonAllworden2018}, and superconducting-tip techniques\cite{Hamidian2016, Liu2021}. These capabilities have enabled atomic-scale studies of unconventional superconductivity\cite{Alldredge2008, Jiao2020, Oh2021}, density-wave states\cite{Hamidian2016, Liu2021, Edkins2019, Du2020, Gu2023}, and topological materials\cite{Avraham2020, Aggarwal2021}. At the same time, modern experiments increasingly generate multidimensional datasets involving spatial coordinates, energy, temperature, magnetic field, and other experimental parameters.

These advances create software requirements beyond numerical processing alone. STM workflows may involve heterogeneous instrument files, intermediate datasets, different image- and curve-based visualization modes, processing-history tracking, and both established and project-specific analysis methods. Existing programs such as WSxM\cite{Horcas2007}, Gwyddion\cite{Necas2012}, ImageJ\cite{Rueden2017}, SpmImage Tycoon\cite{Riss2022}, SPIW\cite{Stirling2013}, and SpectraFox\cite{Ruby2016} address different parts of these requirements. Hystorian\cite{Musy2021} provides a complementary programmatic approach based on standardized HDF5 storage and traceable processing through Python wrappers. AngstromPro targets a different level of the workflow: an interactive STM-oriented application in which on-disk data access, in-application data management, visualization, processing, and extensibility are coordinated within a common GUI architecture.

Here we present AngstromPro, an open-source Python-based software platform for STM data management, visualization, and analysis. Its architecture uses independent module instances with their own Workspaces to organize in-application data, while module-specific central regions provide different visualization and interaction models. Scientific processing methods are registered independently of individual modules and operate on explicitly selected inputs through shared execution services. A common I/O subsystem maps heterogeneous on-disk formats to supported in-application data types, while a plugin subsystem allows modules, processing methods, file handlers, and data types to be extended without editing the core source tree. Representative image-processing and registration workflows are used to demonstrate the resulting architecture and provide quantitative validation of the registration implementation. The contribution of this work lies in the integrated software architecture that coordinates data management, visualization, processing, and extensibility.


\section{Software design}
\begin{figure}
	\centering
	\includegraphics[width=\textwidth,trim={118bp 95bp 110bp 100bp},clip]{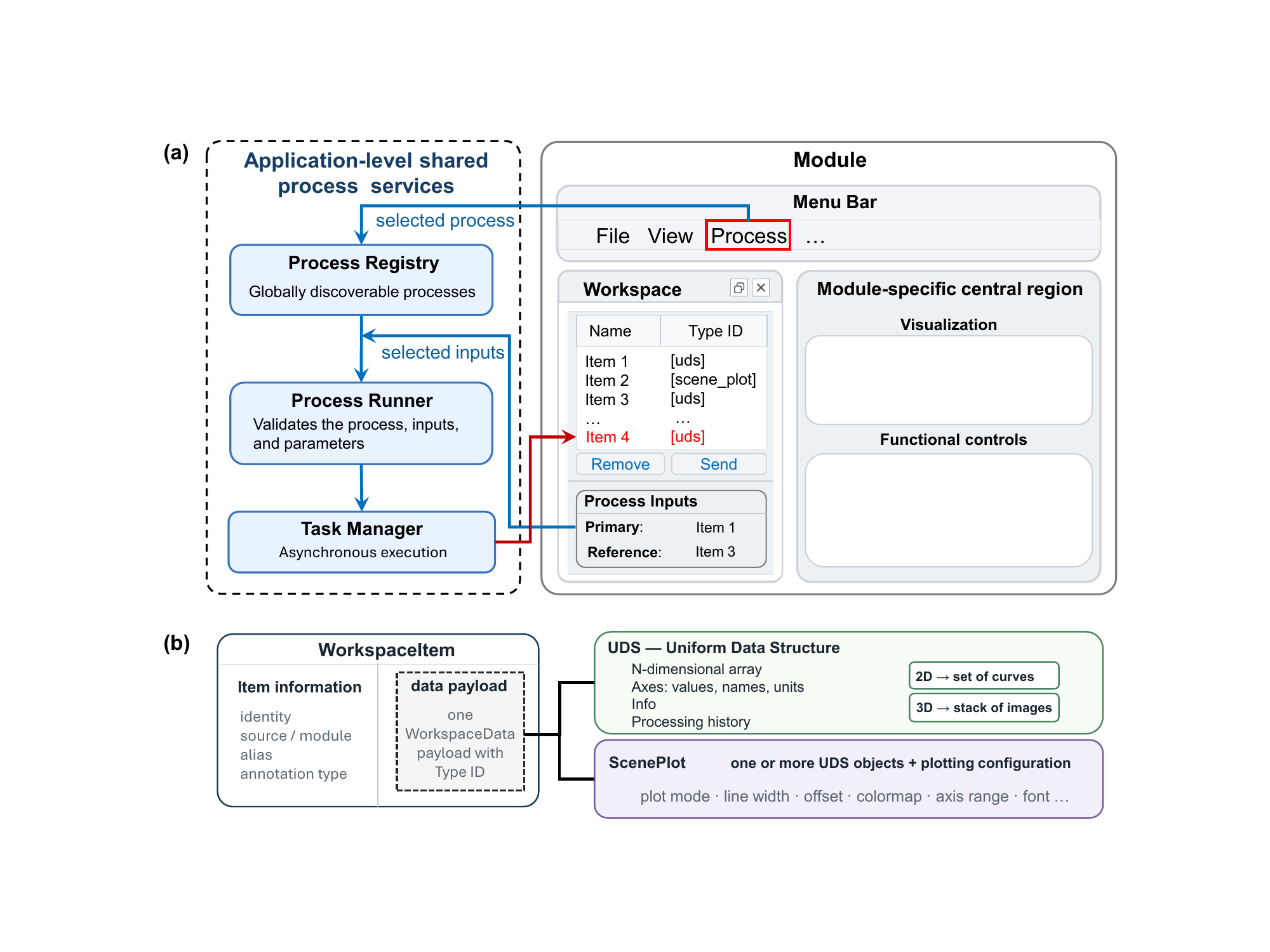}
	\caption{Core application architecture and in-application data model of AngstromPro. (a) Each module instance contains its own Workspace with a common structure for data management and a module-specific central region for visualization or other interactions. Visualization state is separated from the Primary Input and optional Reference Input. Scientific processes are managed independently through the Process Registry, Process Runner, and Task Manager, with processed results returned to the Workspace. (b) WorkspaceItems are the basic units stored in and transferred between module Workspaces. Each WorkspaceItem contains item information and one WorkspaceData payload. The Uniform Data Structure (UDS) provides the basic numerical representation, including an N-dimensional array, coordinate axes, metadata, and processing history; current two- and three-dimensional UDS objects represent sets of curves and stacks of images, respectively. ScenePlot provides a richer representation by combining one or more UDS objects with plotting configuration.}
	\label{figArchitecture}
\end{figure}

AngstromPro separates three basic aspects of scientific data workflows: data management, visualization, and analysis. Figure~\ref{figArchitecture} summarizes the application architecture used to keep these aspects related but distinct. In-application data are organized within independent module instances, presented through module-specific interfaces, and analyzed through registered scientific processes. Throughout this section, we distinguish data stored as files on disk from data objects used within AngstromPro, referring to them as on-disk data and in-application data, respectively.

\subsection{Module and process architecture}

Different scientific workflows can require different visualization and interaction models. Placing all such functions in a single main window would make the interface increasingly complex as the software grows. AngstromPro therefore organizes its application interface into modules, as illustrated in Fig.~\ref{figArchitecture}(a). Each module type is designed around a particular visualization or interaction model, and multiple instances of the same or different module types can run simultaneously.

Each module instance contains a Workspace and a module-specific central region. The Workspace has a common structure across module types and manages the in-application data available to that module instance, whereas the central region defines how those data are visualized or interacted with. This separation keeps in-application data management independent of the specific interface required by a module.

The data being visualized are not necessarily the data to be processed. A module may, for example, display several datasets simultaneously while a processing operation acts on only one or two of them. AngstromPro therefore separates visualization state from the Process Inputs. Each module provides a Primary Input and, when required, an optional Reference Input for operations involving a second dataset. The Process Inputs may coincide with the currently visualized data for certain modules, but this is not assumed by the architecture.

Scientific processes are defined independently of individual modules. Each registered process declares the requirements of its inputs, including the in-application data type and, where relevant, dimensionality or other constraints. A module likewise declares the data requirements of its visualization or interaction interface. Registered processes are globally discoverable and can be assigned to compatible module menus according to these requirements, rather than being hard-coded into individual modules. When a process is invoked, the selected inputs and parameters are passed through the Process Runner to the Task Manager for execution, and the resulting data are returned to the module Workspace. A process is therefore associated with compatible data rather than with a particular GUI module.

\subsection{Workspace and in-application data model}

Independent modules require a common data representation so that scientific data can be stored, transferred, and processed without being tied to a particular visualization interface. At the same time, STM datasets must retain their physical coordinate definitions, while more specialized objects may require information beyond a numerical array. AngstromPro therefore separates a minimal numerical representation from richer application-specific data types, as summarized in Fig.~\ref{figArchitecture}(b). We define the Uniform Data Structure (UDS) as the basic numerical in-application data type in AngstromPro. UDS is intentionally kept minimal. It contains an N-dimensional numerical array together with physical coordinate axes, axis names and units, processing history, and a flexible \texttt{info} field for additional metadata. Basic physical context is retained in the core structure, while dataset- or experiment-specific information can be stored in \texttt{info} when required.

One UDS object represents one basic dataset unit. Different acquisition channels are therefore represented by separate UDS objects. Likewise, datasets are combined only when their coordinate axes---including axis names, values, and units---are identical; otherwise, they are kept as separate UDS objects to avoid ambiguous alignment. For example, two image datasets acquired on different spatial grids, or two spectroscopy datasets sampled at different bias values, are stored separately even if their arrays have the same shape and number of data points. In the current implementation, two-dimensional UDS objects represent sets of curves, while three-dimensional UDS objects represent stacks of images.

UDS is not intended to contain all information required by more specialized data objects. When additional structure is needed, a separate in-application data type can instead be composed from one or more UDS objects together with information specific to its purpose. ScenePlot is one currently implemented example. It combines the underlying numerical data with the plotting configuration required to reproduce a visualization. The basic numerical data type can therefore remain minimal, while richer in-application data types are introduced when needed.

Different in-application data types are managed through a common container, WorkspaceItem. A WorkspaceItem consists of two parts: item-level information and one data payload. The item-level information records properties used for data management within AngstromPro, such as the item identity, source or originating module, alias, and annotation type. The payload derives from the common WorkspaceData base class and carries a Type ID used to identify its data type. UDS and ScenePlot are built-in WorkspaceData types, while additional types can be introduced through plugins. A WorkspaceItem serves as the basic unit stored in a Workspace and transferred between the Workspaces of different module instances. The Workspace itself does not restrict the specific WorkspaceData type carried by an item; compatibility is checked when a WorkspaceItem is used for visualization or supplied to a process, according to the requirements declared by the corresponding module or process.

\subsection{Data I/O and on-disk storage}

\begin{figure}
	\centering
	\includegraphics[width=\textwidth,trim={34bp 186bp 35bp 162bp},clip]{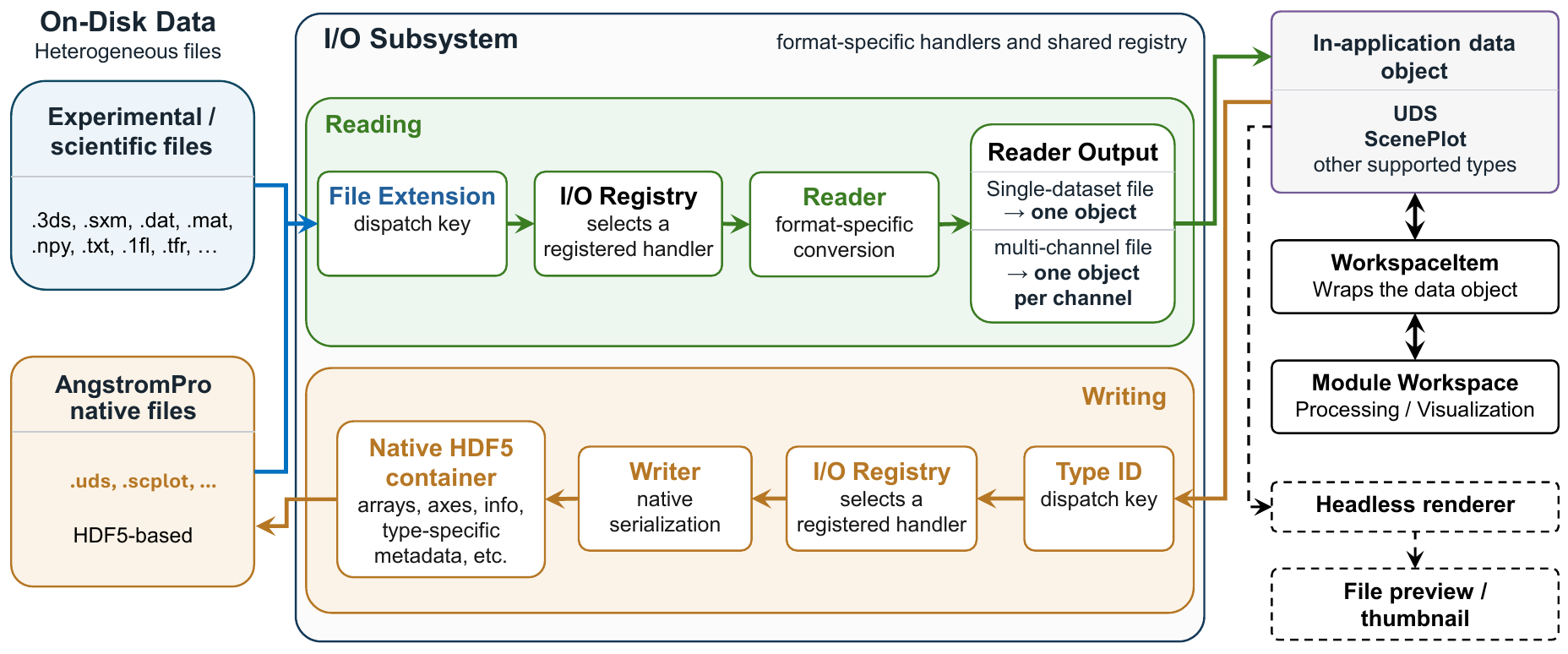}
	\caption{Mapping between on-disk files and in-application data through the AngstromPro I/O subsystem. Both established experimental/scientific formats and AngstromPro native formats are read through registered format-specific handlers selected according to file extension. A Reader may generate one in-application data object or, for multichannel files, one object for each independently usable channel. For writing, the data Type ID is used to select the corresponding Writer, and supported data are stored in AngstromPro native HDF5-based formats. Loaded data can be wrapped in WorkspaceItems for visualization or processing in module Workspaces, or passed to a headless renderer to generate file previews.}
	\label{figIO}
\end{figure}

A common in-application data model is useful only if heterogeneous acquisition files can be converted into it without exposing file-format-specific details to the rest of the application. STM and scientific file formats differ in internal structure, metadata organization, dimensionality, and channel composition. AngstromPro therefore confines format-specific handling to the I/O subsystem, which converts on-disk files into the in-application representations described in Sec.~2.2, as summarized in Fig.~\ref{figIO}.

When reading a file, its extension is used by the I/O Registry to select the corresponding Reader. This applies to both external file formats and AngstromPro native formats. Depending on the file content, a Reader can generate one or multiple in-application data objects. For measurement files containing several independently usable channels, each channel is converted into a separate UDS object. Thus, file-format-specific differences are handled during data loading, and the visualization and processing components operate on the resulting in-application data objects without needing to know the original file format.

The reverse mapping follows a different rule. When saving data, the Writer operates on the WorkspaceData payload contained in a WorkspaceItem rather than on the WorkspaceItem itself. Because an in-application data object does not have a file extension, the appropriate Writer is selected according to its Type ID. Processed data are stored in AngstromPro native formats rather than reconstructed into the original acquisition formats. For example, UDS objects are written in the \texttt{.uds} format, while ScenePlot objects are written in the \texttt{.scplot} format; other WorkspaceData types can define their own native formats. AngstromPro native formats currently use HDF5 as the underlying container. Its hierarchical structure allows numerical arrays, coordinate axes, metadata, processing records, and type-specific information to be stored together in a single file, while providing a common storage mechanism for data types with different internal structures.

Once generated by a Reader, an in-application data object is independent of its source file format. It can be wrapped in a WorkspaceItem and transferred to a module Workspace for visualization or processing. The same representation can also be passed to a renderer to generate a preview of an on-disk file; this function is used by the browsing interface introduced later. Additional file handlers can be incorporated through the I/O system without changing the in-application data model.

\subsection{Built-in modules and application workflow}

Sections~2.1--2.3 describe the shared infrastructure for in-application data management, processing, and I/O. This shared architecture is intended to support scientific activities with genuinely different interaction requirements rather than forcing them into a single universal interface. The built-in modules illustrate how this principle is applied in practice. At startup, AngstromPro instantiates a single Main Workbench, from which other module instances are created as needed. The Main Workbench itself follows the same module architecture, but its central region is used for application-level coordination rather than scientific data visualization. Thus, module management remains within the same architectural framework as the other modules.

\begin{figure}
	\centering
	\includegraphics[width=\textwidth,trim={146bp 39bp 109bp 39bp},clip]{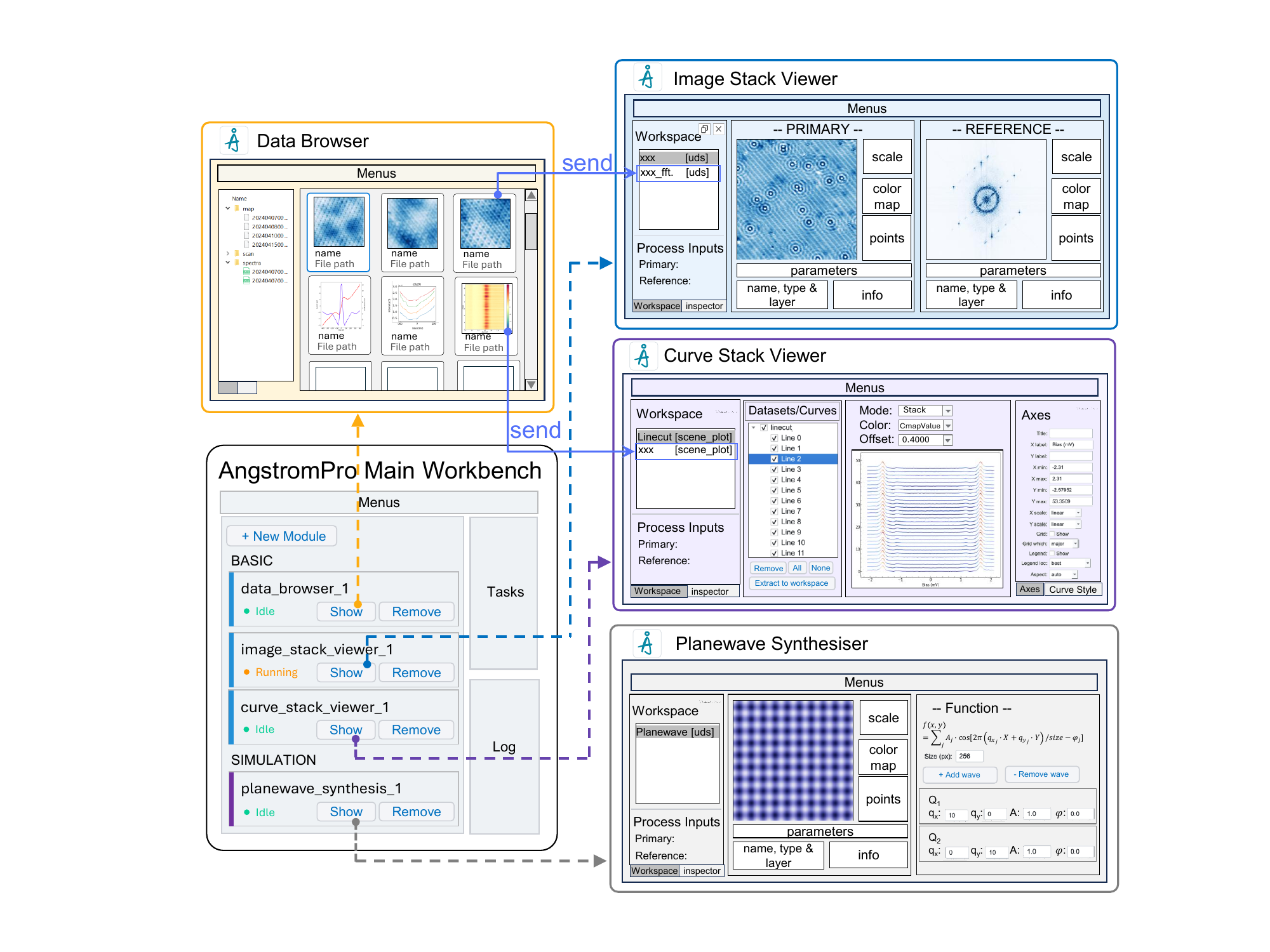}
	\caption{Representative built-in modules and application workflow in AngstromPro. The Main Workbench provides the application entry point from which module instances are created and managed. The Data Browser supports preview-based inspection of on-disk files and transfer of selected data to other modules. The Image Stack Viewer provides Primary and Reference image canvases that correspond to its Primary Input and Reference Input, supporting layer-resolved visualization, real- and reciprocal-space inspection, and comparison of two image datasets. The Curve Stack Viewer displays multiple curves in a common plotting canvas and supports preservation of plotting configuration through ScenePlot data. The Planewave Synthesiser generates synthetic real-space patterns from specified wave vectors, amplitudes, and phases. The connections indicate module management by the Main Workbench and data transfer between module instances.}
	\label{figBuiltinModules}
\end{figure}

Built-in modules implement common activities such as browsing on-disk data, visualizing numerical datasets, and generating synthetic data. Each module instance maintains its own Workspace while using the common Workspace architecture and process infrastructure described above, and its central region is specialized for a particular interaction model. Figure~\ref{figBuiltinModules} shows representative built-in modules and the transfer of WorkspaceItems between their module instances.

Experimental datasets are often distributed across many files and folders, and filenames alone may be insufficient for identifying data of interest. AngstromPro therefore provides the Data Browser for visual inspection of files across multiple user-defined folders. File previews are generated through the same I/O system. To keep browsing responsive, previews are cached and generated asynchronously, with files in the current view given higher priority. These previews are used only for browsing; when a file is selected for further use, the actual data are reloaded through the I/O system and can be sent to another module instance.

The Image Stack Viewer and Curve Stack Viewer provide visualization interfaces for the two common UDS representations introduced in Sec.~2.2. Their central regions reflect the different visualization and interaction requirements of image and curve data.

The Image Stack Viewer operates on three-dimensional UDS objects representing stacks of two-dimensional images. For example, a spectroscopic imaging dataset may contain conductance maps acquired at different energies, with each energy corresponding to one image layer. STM image analysis often requires simultaneous comparison of two such two-dimensional fields, such as a real-space image and its Fourier transform, or conductance maps acquired at different energies. Because superimposing two full images would obscure their spatial information, the central region provides two image canvases for side-by-side inspection. By default, the second canvas displays the Fourier transform of the image shown in the first, while it can alternatively display another dataset for comparison or operations involving two images, such as subtraction or registration. Each canvas supports layer-resolved visualization and spatial interaction. In the Image Stack Viewer, the two canvases are designated Primary and Reference and correspond directly to the Primary Input and Reference Input, respectively, so the displayed datasets can coincide with the Process Inputs; this module-specific correspondence is not imposed by the general architecture described in Sec.~2.1.

The Curve Stack Viewer operates on two-dimensional UDS objects representing collections of curves. Curves stored in different UDS objects may have different coordinate values, ranges, or sampling intervals, but can still be displayed in the same two-dimensional plotting canvas. They can therefore be overlaid or vertically offset for comparison. The central region uses a single canvas that can display curves from multiple WorkspaceItems simultaneously. The displayed curve set remains separate from the Process Inputs, allowing one or a subset of the displayed datasets to be selected for subsequent operations.

Curve visualization can also require substantial plotting configuration beyond the underlying numerical data. When several traces are combined into one figure, axis ranges, offsets, line styles and widths, plotting modes, labels, and other settings may need to be retained together. The Curve Stack Viewer therefore allows the resulting visualization to be stored as a ScenePlot object, so that the numerical data and associated plotting configuration can be reloaded together.

Some STM analyses also benefit from synthetic reference data for examining how selected wave vectors, amplitudes, and phases produce real-space patterns. The Planewave Synthesiser provides this function by generating synthetic datasets from user-defined wave vectors with specified amplitudes and phases; these datasets can also be used to test analysis procedures.

\subsection{Plugin subsystem and extensibility}

\begin{figure}
	\centering
	\includegraphics[width=\textwidth,trim={34bp 164bp 55bp 137bp},clip]{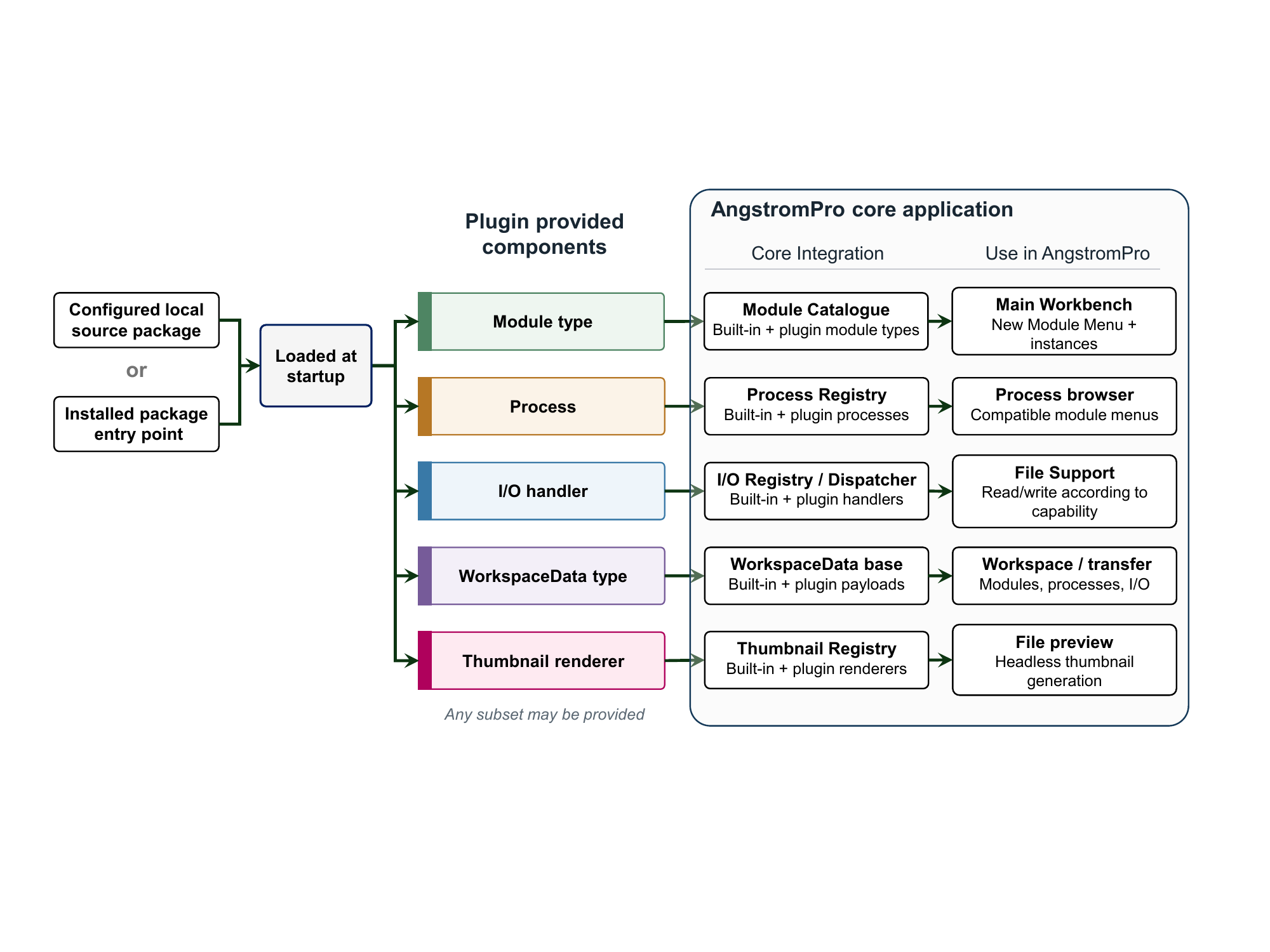}
	\caption{Plugin loading and extension paths in AngstromPro. External plugins are loaded at application startup from either configured local source packages or installed package entry points. A plugin may provide any subset of module types, scientific processes, I/O handlers, WorkspaceData based payload types, and optional thumbnail renderers. These components are integrated through the same Module Catalogue, Process Registry, I/O Registry / Dispatcher, WorkspaceData base class, and Thumbnail Registry used by built-in functionality, making them available to the Main Workbench, compatible module menus, file I/O, Workspace transfer, and file-preview mechanisms without editing the AngstromPro source tree.}
	\label{figPluginSubsystem}
\end{figure}

The built-in modules, processes, and file handlers cannot cover every experimental workflow, analysis method, or instrument format. Incorporating every future extension directly into the core package would also increase coupling and make the architecture progressively harder to maintain. AngstromPro therefore exposes the same registration interfaces used internally to external plugins, allowing new functionality to be added without modifying the core source tree. Plugins are discovered at application startup from user-specified or installed Python packages and can register modules, processes, I/O handlers, WorkspaceData types, and thumbnail renderers, as summarized in Fig.~\ref{figPluginSubsystem}.

The extension points are independent, so a plugin may provide any subset of these components. Registered modules become available through the Main Workbench. Processes become globally discoverable and can be assigned to compatible module menus according to their declared input requirements. I/O extensions can provide either native read/write support for a WorkspaceData type or a read-only loader for an external file format.

Plugins can also introduce new WorkspaceData types when the built-in representations described in Sec.~2.2 are insufficient. These WorkspaceData types can use the same Workspace storage and transfer mechanisms as built-in types and can participate in the registered process and I/O infrastructure. An optional thumbnail renderer can also be provided when preview support is required. In this way, new data representations can be added without expanding the basic UDS definition or changing the Workspace model.

The example plugin distributed with AngstromPro registers a GUI module, a scientific process, and a read-only file loader, demonstrating module, process, and external-file I/O extensions without modification of the core package. Detailed plugin interfaces and example code are provided in the project documentation.

\section{Representative workflow and validation}

AngstromPro currently provides a set of built-in processing methods, most of which implement well-established STM and digital image-processing approaches\cite{Burger2022}. Additional methods, including newly developed algorithms, can be incorporated through the same registration and plugin mechanisms. The present work focuses on the design and validation of the software platform rather than on the novelty of individual algorithms. Accordingly, this section uses representative well-established image-processing and registration methods to illustrate how processing methods are integrated into AngstromPro and to provide quantitative validation of the registration implementation.

\subsection{Representative image-processing workflow}

\begin{figure}
	\centering
	\includegraphics[width=13.5cm]{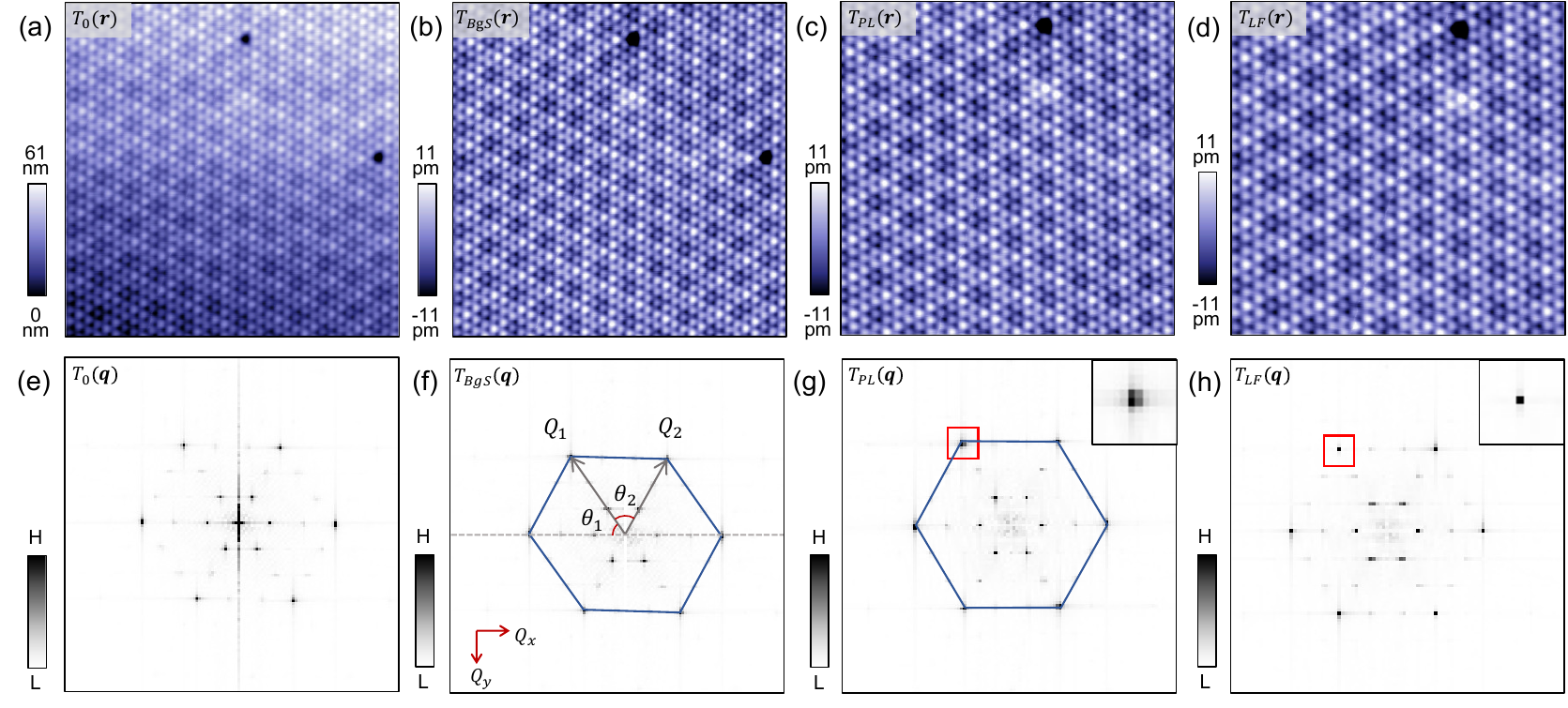}
	\caption{Representative STM image-processing workflow. (a) Raw NbSe$_2$ topography $T_0(\mathbf{r})$. (b--d) Results after sequential background subtraction, perfect-lattice (affine) correction, and Lawler--Fujita drift correction, respectively. (e--h) Fourier transforms of (a--d), respectively. In (f), $\mathbf{Q}_1$ and $\mathbf{Q}_2$ denote the two reciprocal-lattice vectors; $\theta_1$ is the angle between $\mathbf{Q}_1$ and the horizontal axis, and $\theta_2$ is the angle between $\mathbf{Q}_1$ and $\mathbf{Q}_2$. The insets in (g) and (h) enlarge the red-boxed Bragg-peak regions and illustrate the peak sharpening after Lawler--Fujita correction.}
	\label{figBgPlLf}
\end{figure}

Figure~\ref{figBgPlLf} presents a representative STM image-processing workflow using a NbSe$_2$ topographic dataset. The dataset is first inspected in the Data Browser and transferred to the Image Stack Viewer, where the real-space topography and its Fourier transform are displayed together. Background subtraction, perfect-lattice (affine) correction, and Lawler--Fujita drift correction are then applied sequentially using the corresponding built-in processing methods. Each processing step generates a new data object in the module Workspace, allowing the intermediate results and their Fourier transforms to be inspected and compared throughout the workflow.

These well-established methods are used here to demonstrate how multiple processing steps are integrated within AngstromPro rather than to introduce the methods themselves. In the example shown, background variation is first removed, the reciprocal-lattice geometry is then corrected, and the subsequent Lawler--Fujita correction sharpens the Bragg peaks by reducing residual spatial distortion.

In AngstromPro, the Gaussian-windowed 2D lock-in operation used in the Lawler--Fujita correction\cite{Lawler2010} is evaluated as a convolution in Fourier space using the convolution theorem, rather than by direct real-space integration at every image position. For an $N \times N$ image, a naive direct evaluation requires approximately $\mathcal{O}(N^4)$ operations, whereas the FFT-based implementation scales approximately as $\mathcal{O}(N^2 \log N)$. This implementation preserves the established method while reducing the computational cost of the local lock-in evaluation.

\subsection{Quantitative validation by image registration}

Image registration\cite{Burger2022} provides a convenient quantitative test of the implementation because the spatial agreement between two images can be evaluated independently after the transformation. In this example, two consecutively acquired STM topographies are loaded into the Primary Input and Reference Input of the Image Stack Viewer. Synchronized zooming and panning of the two displayed datasets facilitates the identification of three pairs of corresponding landmarks, which are then used to determine the affine transformation.

\begin{figure}
	\centering
	\includegraphics[width=13.5cm,trim={177bp 163bp 207bp 203bp},clip]{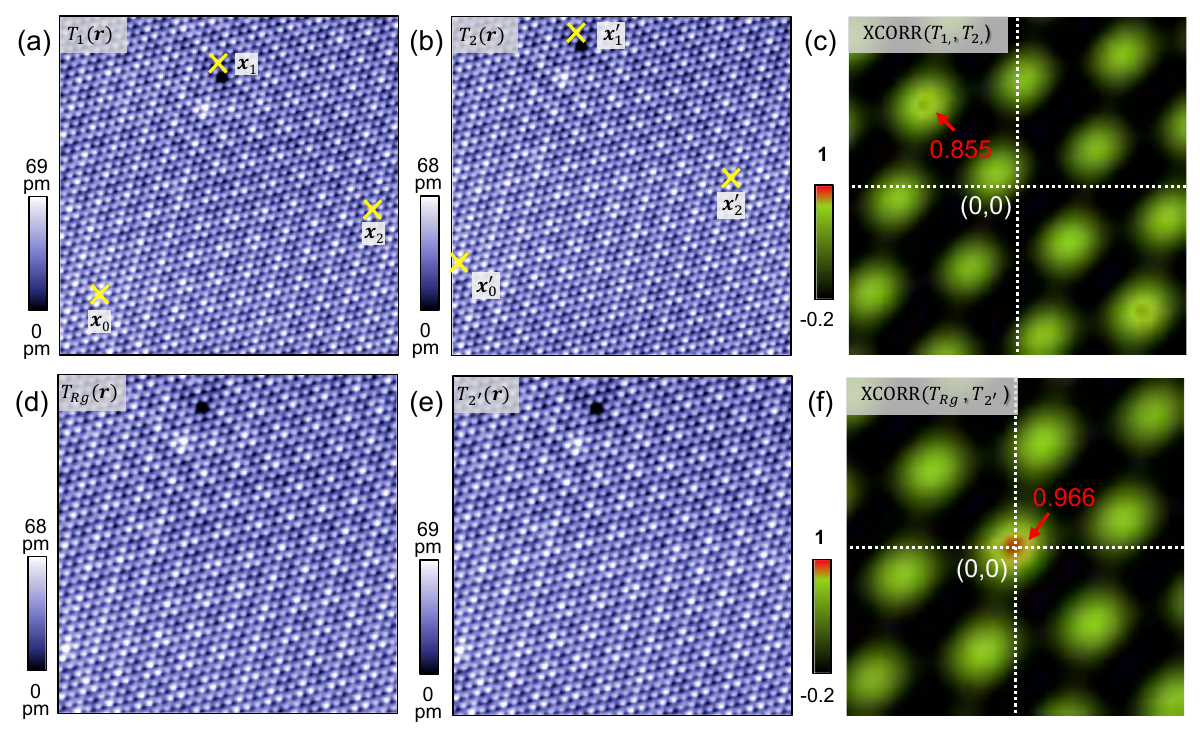}
	\caption{Image-registration workflow and quantitative validation. (a,b) Two consecutively acquired STM topographies, $T_1(\mathbf{r})$ and $T_2(\mathbf{r})$, with three pairs of corresponding landmarks $x_0,x_1,x_2$ and $x'_0,x'_1,x'_2$ marked by yellow crosses. (c) Normalized cross-correlation of (a) and (b). The dashed crosshairs indicate zero relative displacement, and the red arrow marks the off-center correlation maximum of 0.855. (d) Registered and cropped image obtained from (a), $T_{\mathrm{Rg}}(\mathbf{r})$. (e) Correspondingly cropped reference image from (b), $T'_2(\mathbf{r})$. (f) Normalized cross-correlation of (d) and (e), for which the maximum increases to 0.966 and moves close to zero relative displacement.}
	\label{figRegister}
\end{figure}

Figure~\ref{figRegister} compares the two topographies before and after registration. Their spatial agreement is evaluated by normalized cross-correlation as a function of relative displacement. Before registration, the correlation maximum is displaced from zero relative displacement and has a value of 0.855 [Fig.~\ref{figRegister}(c)]. After registration, the maximum coincides closely with the zero-displacement position and increases to 0.966 [Fig.~\ref{figRegister}(f)]. The shift of the correlation maximum toward $(0,0)$, together with the increased peak value, indicates improved spatial alignment between the two images. This example provides a quantitative validation of the registration implementation in AngstromPro while demonstrating how an established image-processing method is incorporated into the platform workflow.

\section{Conclusion}
We have presented AngstromPro, an open-source Python-based software platform for STM data management, visualization, and analysis. The software separates in-application data management, module-specific visualization, and scientific processing while connecting them through common Workspace, process-execution, and I/O infrastructures. Independent module instances support different scientific working contexts, and the plugin subsystem allows additional modules, processing methods, file handlers, and data types to be incorporated without modifying the core package. Representative STM image-processing and registration workflows were used to demonstrate how established analysis methods are integrated into this architecture. The registration example further provides quantitative validation through the increase in the normalized cross-correlation maximum from 0.855 to 0.966 and the shift of its peak toward zero relative displacement after alignment. The principal contribution of AngstromPro is an integrated software architecture that brings data management, visualization, scientific processing, and extensibility into a common interactive environment for STM data analysis.

\printcredits

\section*{Data and code availability}
The data that support the findings of this study are available from the corresponding author upon reasonable request. The source code is openly available on GitHub at \url{https://github.com/jhyan2018/AngstromPro}.

\section*{Acknowledgments}
We acknowledge the influence of software privately developed by Kazuhiro Fujita, which inspired elements of this work. We thank J. C. Séamus Davis, Ge He, and Siyuan Wan for discussions and advice. This work was supported by grants from the National Key Research and Development Projects of China (2022YFA1204100), the National Natural Science Foundation of China (62488201), the Innovation Program of Quantum Science and Technology (2021ZD0302700), and Research Ireland (17/RP/5445).

\bibliographystyle{elsarticle-num}
\bibliography{refs}

\clearpage

\clearpage



\end{document}